\pdfoutput=1
\documentclass[aps,prl,twocolumn,superscriptaddress,floatfix]{revtex4-2}

\usepackage{graphicx}
\usepackage{amsmath,amssymb}
\usepackage{booktabs}
\usepackage{xcolor}
\usepackage[colorlinks=true,linkcolor=blue!55!black,citecolor=blue!55!black,urlcolor=blue!55!black]{hyperref}

\newcommand{\Fxeb}{F_{\mathrm{XEB}}}
\newcommand{\Camp}{C_{\mathrm{amp}}}
\newcommand{\panel}[1]{\textbf{(#1)}}

\begin{document}
\setcounter{secnumdepth}{3}

\title{Quantum computational advantage in random-circuit sampling on IBM superconducting quantum computers}

\author{Tigran Sedrakyan}
\affiliation{BlueQubit, San Francisco, CA 94105, USA}
\author{Yuxuan Zhang}
\affiliation{BlueQubit, San Francisco, CA 94105, USA}
\affiliation{Institute of Physics, \'Ecole Polytechnique F\'ed\'erale de Lausanne, Lausanne, Switzerland}
\author{Hovnatan Karapetyan}
\affiliation{BlueQubit, San Francisco, CA 94105, USA}
\author{Joshua D. Baktay}
\affiliation{BlueQubit, San Francisco, CA 94105, USA}
\affiliation{XPRIZE Foundation, Culver City, CA 90230, USA}
\author{Hrant Gharibyan}
\affiliation{BlueQubit, San Francisco, CA 94105, USA}
\author{Hayk Tepanyan}
\affiliation{BlueQubit, San Francisco, CA 94105, USA}

\date{\today}

\begin{abstract}
We report forward random-circuit sampling (RCS) on the 120-qubit Nighthawk r2 superconducting processor (\textit{ibm\_phoenix}) with square-lattice connectivity, using 78 qubits, native CZ gates, and the standard cloud execution stack with no benchmark-specific calibration. The independent fidelity estimators---mirror benchmarking and three- and four-patch cross-entropy benchmarking (XEB)---agree with each other at every measured depth, across more than two orders of magnitude of fidelity decay. The deepest 40-cycle circuits sit at the depth where tensor-network contraction cost saturates at system size: a contraction-cost estimator validated against the previously published Sycamore and Zuchongzhi networks places the single-amplitude cost at $\sim$$10^{27.4}$ complex operations. At $\Fxeb(40)=1.8\times10^{-4}$ under favorable memory assumptions this implies $3.5\times10^{31}$ machine operations within the bounded-fidelity rejection-sampling model, and even an unlimited-memory state-vector simulation would take $1.2\times10^{3}$ years --- more than a millennium of runtime on the Frontier supercomputer --- to collect a $10^{6}$-sample ensemble, which takes only 22\,s on Nighthawk r2. To our knowledge, this is the first demonstration of quantum advantage for a vanilla random-circuit sampling on a commercially and broadly accessible quantum processor that most non-expert quantum computer users can easily replicate.
\end{abstract}

\maketitle

\section{Introduction}
The computational power of a quantum processor derives from the exponential growth of Hilbert-space dimension with qubit number; on near-term devices it is bounded instead by how much of that space noise leaves coherently accessible. Random-circuit sampling (RCS), scored by linear cross-entropy benchmarking (XEB), has become the standard probe of this boundary: quantum correlations spread as fast as the geometry allows, and hardness rests on anticoncentration and average-case conjectures for the output probabilities; what the experiment actually reports is one number, a fidelity proxy~\cite{preskill2018,aaronsonchen2017,boixo2018,neill2018,bouland2019,dalzell2022,movassagh2023,hangleiter2023}. Since the 53-qubit Sycamore experiment~\cite{arute2019}, the quantum side of this competition has advanced through Zuchongzhi~2.0 and~2.1~\cite{wu2021,zhu2022}, Sycamore at 67 qubits~\cite{morvan2024}, and Zuchongzhi~3.0 at 83 active qubits and 32 cycles~\cite{gao2025}, while the classical side has cut the cost of Sycamore-class verification and sampling by orders of magnitude~\cite{boixo2017,chen2018,villalonga2019,huang2020,liu2021,gray2021,kalachev2021,pan2022,liu2024,zhao2025}. Trapped-ion processors with all-to-all connectivity have meanwhile run RCS at 56 and 98 qubits~\cite{decross2025,ransford2026}, estimating fidelity by mirror benchmarking. The competition has also sharpened what such claims mean. XEB certifies no total-variation closeness: it can be spoofed in exploitable regimes, and it tracks fidelity only in a limited window of depth and noise~\cite{morvan2024,aaronsongunn2020,barak2021,gao2024,ware2023}. Hardness hinges on three crossovers that are easy to conflate: the depth at which outputs anticoncentrate, the noise past which subsystem structure takes over the score, and the depth past which contraction cost saturates at system size~\cite{dalzell2022,morvan2024,gray2021,ware2023,markov2008}, and under specified local-noise models, noisy anticoncentrated RCS can be sampled classically outright~\cite{bouland2021,aharonov2023,dalzell2024,zlokapa2023}. A hardness claim therefore has to say where the experiment sits relative to each of these crossovers, and which classical task the quoted cost refers to; we do both explicitly below.

Here we report forward RCS on the Nighthawk r2 processor
(\textit{ibm\_phoenix}) using 78 qubits of a square-lattice coupler graph, native CZ entangling gates, and a four-color brickwork schedule,\footnote{One cycle applies one of the four coupler colors (A,B,C,D) together with a layer of random single-qubit rotations; four cycles constitute one full sweep of the 133 couplers. Cycle count therefore equals two-qubit depth.} compiled and executed through the standard Qiskit cloud workflow~\cite{qiskit2024} with no benchmark-specific calibration. Fidelity is estimated by mirror benchmarking~\cite{mayer2021,proctor2022} and by three- and four-patch XEB circuits~\cite{arute2019,morvan2024,gao2025}, which all agree with each other at every measured depth from 20 to 40 cycles once each patch is normalized by its ideal XEB (Sec.~\ref{sec:methods}), and together they cover more than two orders of magnitude of decay in fidelity. At 40 cycles, where the proxy estimators give $F\approx1.8\times10^{-4}$, we collect $M=10^{6}$ samples of the full, unpatched circuit in each of two executions. Direct verification of this set is out of reach---each ideal amplitude costs $\sim10^{27}$ operations, i.e.\ about $1.5\times10^{3}$ years of the Frontier supercomputer, for $10^{6}$ amplitudes (Sec.~\ref{sec:classical})---so the fidelity of the advantage configuration is taken from the two proxy estimators measured at the same depth, with no depth extrapolation. Earlier runs of this protocol on the first-generation Nighthawk r1 device (\textit{ibm\_miami}) are documented, together with their data, in the companion repository.

For local two-dimensional circuits, light-cone and tensor-network arguments predict a schedule-dependent crossover $d^{\ast}=O(\sqrt{n})$ at which contraction cost becomes system-size limited~\cite{morvan2024,boixo2017,gray2021,markov2008}; for our schedule the heuristic estimate $d^{\ast}\simeq4\sqrt{n}$ (the coefficient is schedule-specific) gives $d^{\ast}\simeq35.3$ at $n=78$. Our optimized contraction paths land on this knee: using a hyperoptimized contractor~\cite{gray2021} validated against the published Sycamore-67 and Zuchongzhi-3.0 tensor networks~\cite{morvan2024,gao2025}, the best-found single-amplitude cost rises by roughly $10^{3.5}$ per four cycles up to depth 24 and $10^{2.5}$ from there to depth 36---reaching $C_{\mathrm{amp}}\approx10^{27.2}$ complex operations at contraction width 82---and then changes by a factor of 1.4 between depths 36 and 40. Within the bounded-fidelity rejection-sampling model~\cite{morvan2024,kalachev2021}, the fitted $\Fxeb(40)=1.8\times10^{-4}$ gives $3.5\times10^{31}$ machine operations for the $10^{6}$-sample ensemble---roughly $2.7\times10^{6}$ Frontier years at 20\% of theoretical peak~\cite{atchley2023,top500}; even a state-vector simulation with unlimited memory, which yields every amplitude exactly, would take $1.2\times10^{3}$ Frontier years. We treat this as an optimistic baseline computed under assumptions that favor the classical side (Sec.~\ref{sec:classical}) with unbounded memory and negligible communication \cite{liu2024,decross2025,graychan2024,ayral2023}.

\begin{figure*}[!t]
  \centering
  \begin{minipage}[t]{0.38\textwidth}
    {\small\panel{a}}\\[-1pt]
    \includegraphics[width=\linewidth]{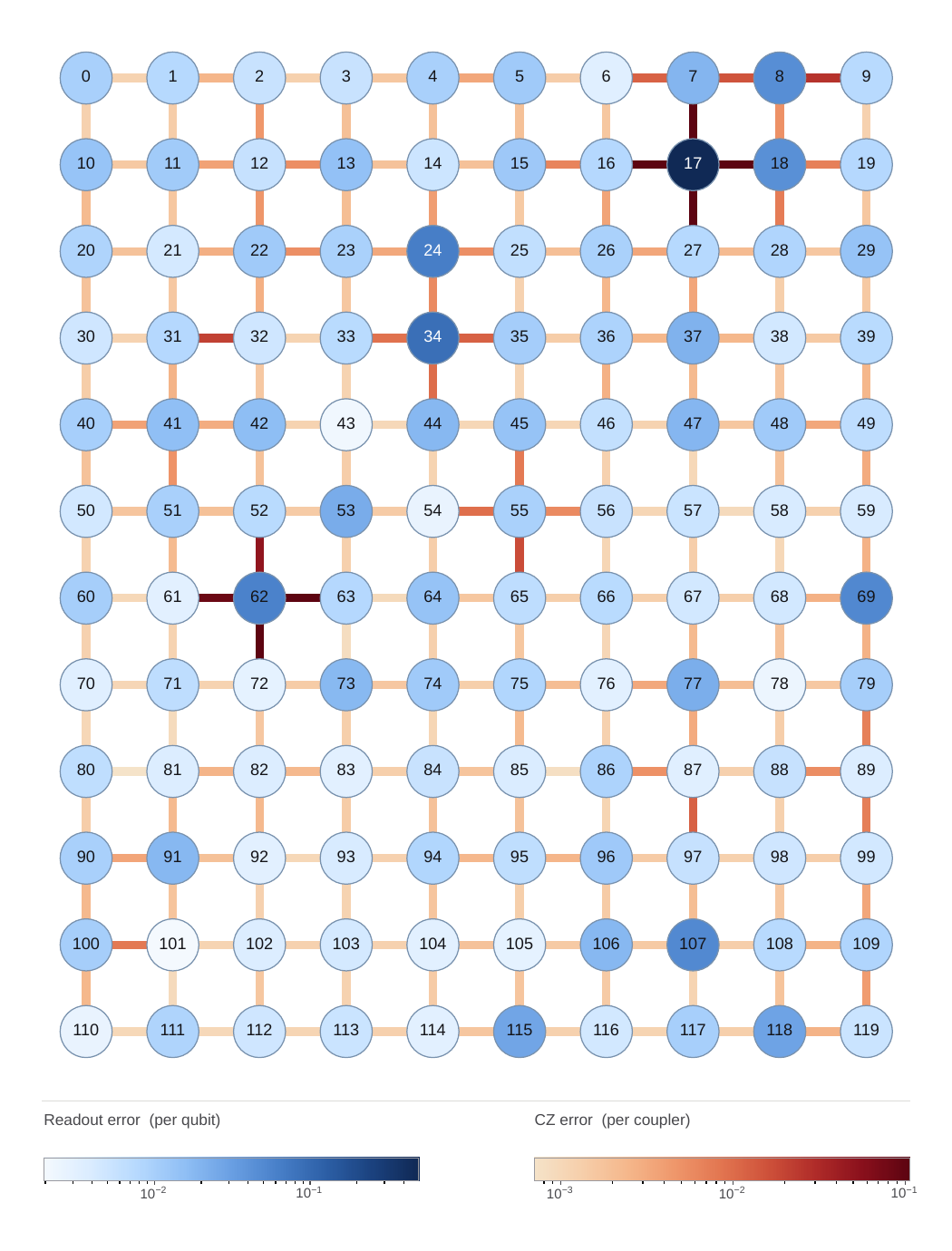}
  \end{minipage}\hspace{0.02\textwidth}%
  \begin{minipage}[t]{0.38\textwidth}
    {\small\panel{b}}\\[-1pt]
    \includegraphics[width=\linewidth]{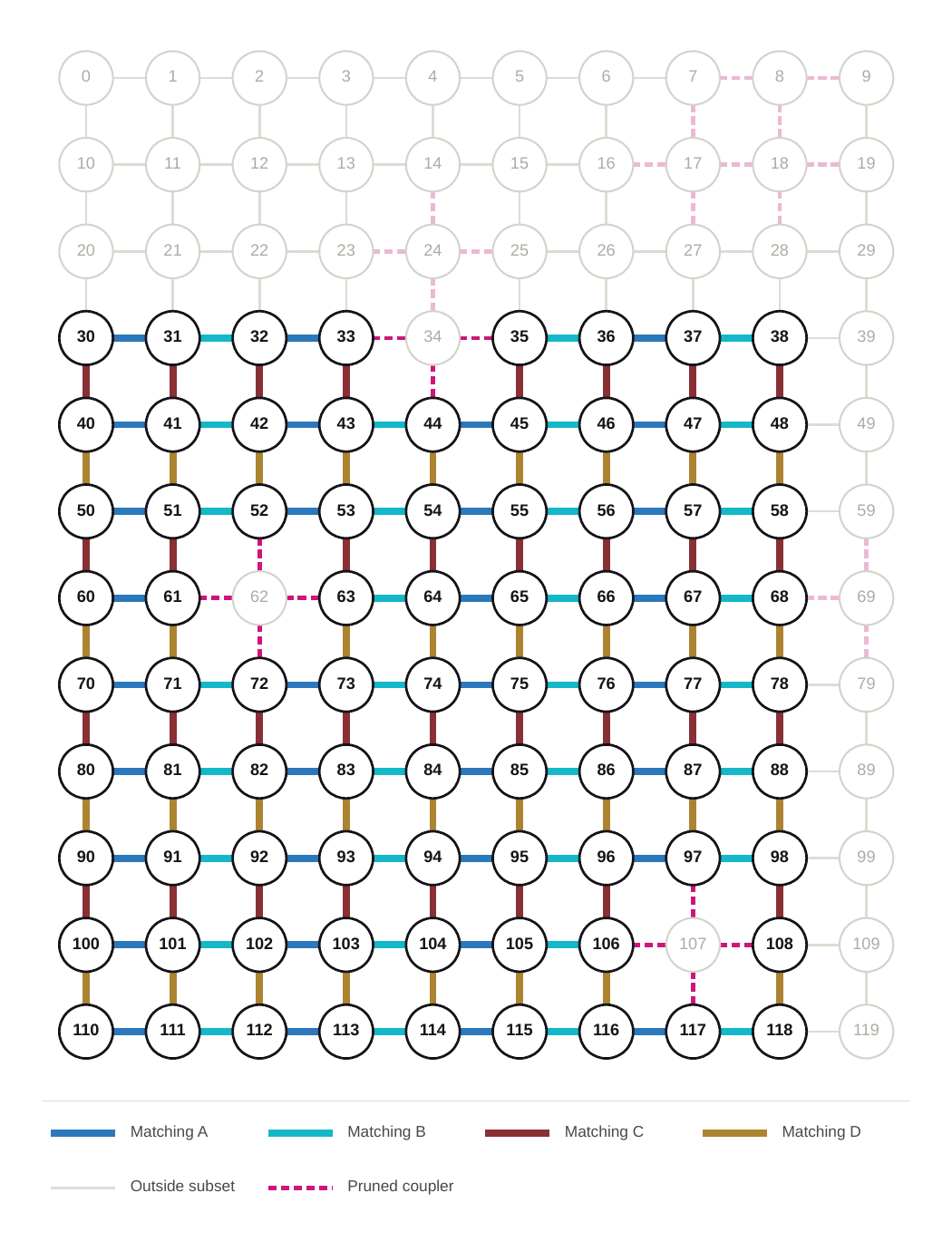}
  \end{minipage}\hspace{0.02\textwidth}%
  \begin{minipage}[t]{0.18\textwidth}
    {\small\panel{c}}\\[-1pt]
    \hspace{1pt}
    \includegraphics[width=\linewidth]{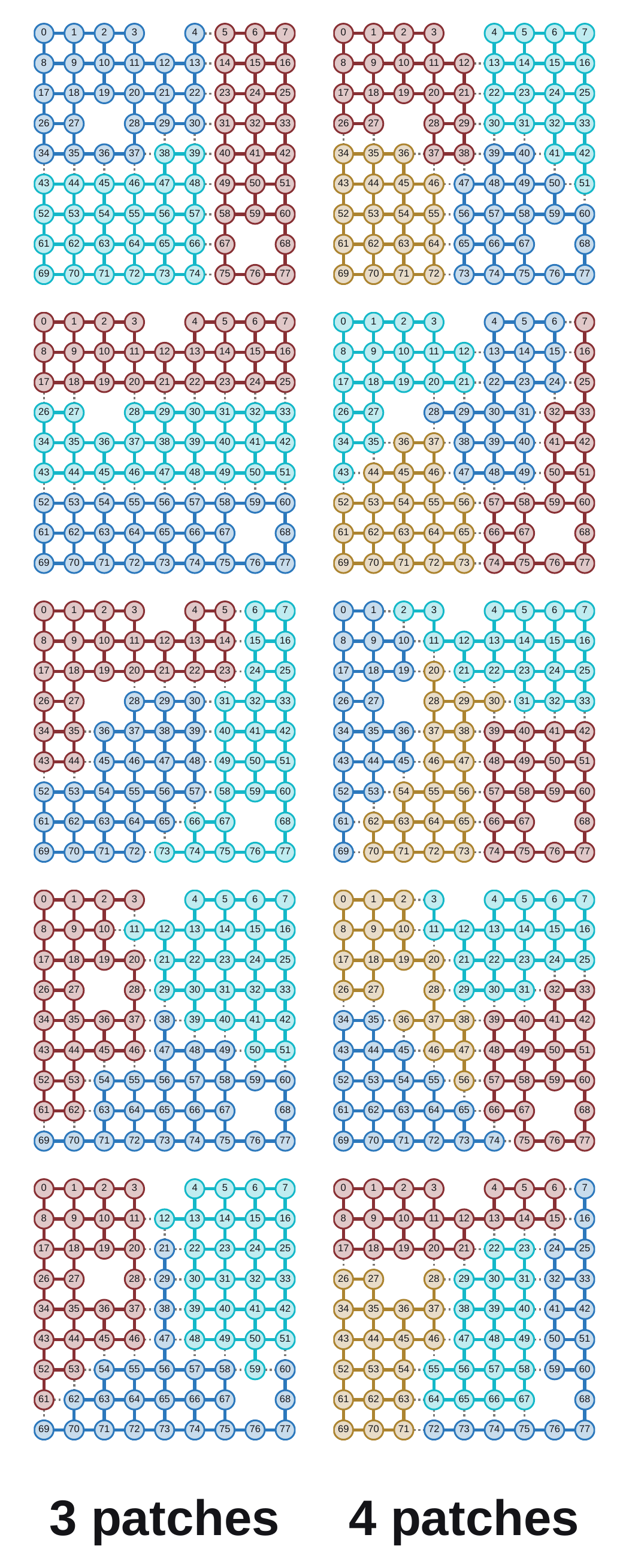}
  \end{minipage}\hfill
  \caption{\textbf{Device and qubit placement on IBM Nighthawk r2.}
  \panel{a}~Calibration snapshot of the 120-qubit square lattice \textit{ibm\_phoenix} device at the time of the experiment: node color gives the qubit readout assignment error, edge color the CZ error.
  \panel{b}~Calibration-aware placement (Appendix~\ref{app:placement}): the selected $9\times9$ subgrid retains $n=78$ qubits and 133 couplers; dropped qubits are grayed out, retained couplers are colored by the four-coloring $\mathrm{ABCD}$ of the entangling schedule, and pruned couplers are dashed.
  \panel{c}~The five randomized patching partitions per patch number ($K=3,4$) used for fidelity estimation (Sec.~\ref{sec:methods}); qubits of each patch share a color, and boundary couplers removed by the partition are dashed.}
  \label{fig:device}
\end{figure*}

\section{Experimental setup and methods}
\label{sec:methods}

\emph{Hardware.}---All experiments were performed on a 120-qubit superconducting QPU based on the Nighthawk r2 architecture. In contrast to the heavy-hexagon connectivity of the preceding IBM chips, Nighthawk arranges its qubits on a $12\times 10$ square lattice whose nearest-neighbor pairs are connected by tunable couplers, i.e.\ the full square-lattice connectivity is available with a total of 218 couplers. This connectivity makes it a natural fit for RCS experiments, which are traditionally performed on two-dimensional grids, allowing the circuit to be laid out on Nighthawk without any routing overhead. The native gate set consists of the two-qubit CZ gate together with one-qubit gates $\mathrm{SX}, \mathrm{X}$ and $\mathrm{RZ}(\theta)$. $\mathrm{RZ}$ rotations are implemented virtually as frame updates and are therefore error-free and of zero duration~\cite{mckay2017efficient}. The device also exposes $\mathrm{RX}(\theta)$ gates, but we do not use them in our experiments, so every one-qubit gate compiles to two SX pulses and virtual RZs. At the time of our runs the device showed a median CZ error of $\sim\!1.5\times 10^{-3}$, median SX error of around $2.1\times10^{-4}$ and a median readout assignment error of $\sim\!6.5\times10^{-3}$. An error map displaying the CZ and readout errors immediately before the experiment is shown in Fig.~\ref{fig:device}(a). Because single-qubit errors are an order of magnitude below the CZ error and RZ is exact, the error budget of the circuits described below is dominated by the two-qubit layers and by readout, which motivates several of the methodological choices that follow.

\emph{Calibration-aware placement.}---Rather than fixing the physical qubit subset by hand, we select it programmatically from the most recent calibration data immediately before each experiment. The target layout is a rectangular $9\times 9$ grid; every rigid offset of the logical grid inside the $12\times 10$ physical lattice is scored by a log-fidelity proxy built from the reported readout, SX, and CZ error rates, with qubits or couplers failing per-element error thresholds dropped from the candidate (Appendix~\ref{app:placement}). Among all candidates within a small slack of the optimal score, the one retaining the most qubits is selected. For the experiment described here, the search retained $n = 78$ qubits (three qubits were dropped by the calibration filters) and 133 couplers between retained qubits, as shown in Fig.~\ref{fig:device}(b). All circuits below act on these $n=78$ qubits, and all gate layers are restricted to the 133 couplers that appear in the executed circuits.

\emph{Random circuit ensemble.}---Our circuits follow the standard RCS construction of alternating one- and two-qubit layers~\cite{boixo2018,arute2019}. One cycle consists of (i) a layer of independent Haar-random SU(2) gates on every qubit, followed by (ii) a layer of CZ gates applied on a matching of the coupling graph. Haar-random one-qubit gates are sampled in the $\mathrm{RZ}(\phi)\,\mathrm{RX}(\theta)\,\mathrm{RZ}(\lambda)$ decomposition with $\phi,\lambda \sim U[0,2\pi)$ and $\cos\theta \sim U[-1,1]$. The two-qubit layers cycle through the canonical four-coloring of the square-lattice edges, $\mathrm{A},\mathrm{B},\mathrm{C},\mathrm{D}$ = (horizontal-even, horizontal-odd, vertical-even, vertical-odd), in the fixed repeating order $\mathrm{ABCD}$. After the placement-driven pruning above the four matchings contain $33/33/33/34$ edges, i.e.\ $33.25$ CZ gates per cycle on average (1330 CZ gates at the deepest benchmarked point, $d=40$ cycles). A depth-$d$ circuit applies $d$ cycles and a terminal measurement of all qubits in the computational basis. Once built, the circuits are transpiled to the native gate set with a locked one-to-one layout onto the qubits selected above, without additional routing.

\begin{figure*}[!t]
  \centering
    \includegraphics[width=0.8\linewidth]{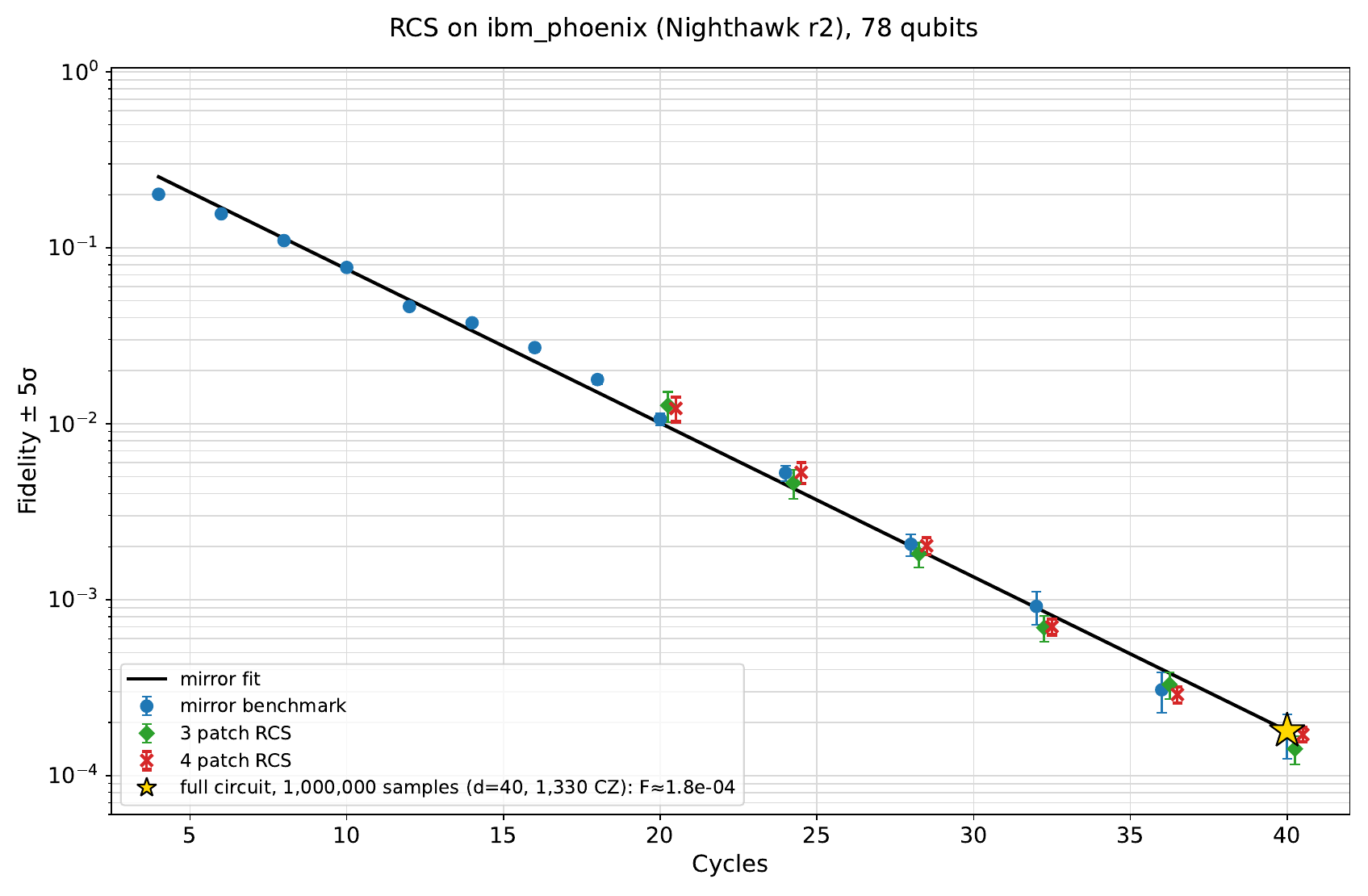}
  \caption{\textbf{Fidelity estimation and main result.}
  Circuit fidelity versus depth on Nighthawk r2 (\textit{ibm\_phoenix}, 78 qubits). Mirror-benchmark estimates (blue, 4--40 cycles) and three- and four-patch XEB estimates (green, red; 20--40 cycles, each patch normalized by its ideal XEB, Eq.~\eqref{eq:patched-xeb-r}) agree with a single exponential fit to the mirror data (solid line, $0.818$ per cycle). The star marks the 40-cycle full circuit (1330 CZ gates) from which $10^{6}$ samples were collected in each of two executions, at the fitted $\Fxeb\approx1.8\times10^{-4}$. Error bars are $\pm5\sigma$, accounting for shot noise.}
  \label{fig:main}
\end{figure*}

\emph{Patched circuits.}---A direct verification through XEB is classically demanding at the scale we probe (Sec.~\ref{sec:classical}). Following the patch-verification methodology introduced previously~\cite{arute2019,morvan2024}, we estimate the fidelity of circuits too large for full classical verification by running patched variants: the qubit set is partitioned into $K$ patches and all CZ gates crossing patch boundaries are removed, so the circuit factorizes into $K$ independent, classically simulable blocks. We benchmark patched families with $K=3,4$.

A single fixed partition would tie the benchmark to one specific set of removed couplers and one specific set of subcircuit ensembles. To separate the fidelity estimate from any particular cut, we use randomized patching [Fig.~\ref{fig:device}(c)]: for every $K$ we take the five best partitions found by a cut-minimizing combinatorial optimizer (Appendix~\ref{app:patchgen}) and allocate three random circuit instances to each, giving 15 independent circuits per data point. Instance-to-instance fluctuations then average over both the random gates and the choice of cut.

For each patched circuit we compute the XEB patch-by-patch~\cite{boixo2018,arute2019}. Let $x^{(s)}$, $s=1\dots S$, be the measured bitstrings, $x^{(s)}_r$ their substrings corresponding to patch $r$ of $n_r$ qubits, and $p_r$ the ideal output distribution of patch $r$ (obtained by exact statevector simulation of the patch subcircuit). The per-patch estimator is then
\begin{equation}
\hat F_r \;=\; \frac{\displaystyle \frac{2^{n_r}}{S}\sum_{s=1}^{S} p_r\!\bigl(x^{(s)}_r\bigr) - 1}{\displaystyle 2^{n_r}\sum_{x} p_r(x)^{2} - 1},
\label{eq:patched-xeb-r}
\end{equation}
where the denominator is the noiseless value of the numerator, evaluated from the same statevector~\cite{arute2019}. For an anticoncentrated patch it equals $(2^{n_r}-1)/(2^{n_r}+1)\approx1$ and Eq.~\eqref{eq:patched-xeb-r} reduces to the standard linear XEB; at shallower depth, where the patch output has not yet anticoncentrated, the normalization removes the corresponding upward bias (for our experiments patched circuits start to anticoncentrate around 24 cycles, see Appendix~\ref{app:spoofing}). The total circuit fidelity is estimated as the product over patches
\begin{equation}
\hat F \;=\; \prod_{r=1}^{K} \hat F_r.
\label{eq:patched-xeb}
\end{equation}
Ideal patch probabilities were computed with statevector simulation on the BlueQubit platform (CPU; patches of at most 26 qubits)~\cite{mazumder2026benchmarkingzerosetupquantumcircuit}.

\emph{Mirror benchmark.}---Patched XEB certifies the constituent patches but is, by construction, blind to errors on the removed boundary couplers, and any XEB-type estimate at these sizes relies on classical simulability of the reference distribution. As an independent, simulation-free consistency check we run a mirror benchmark~\cite{proctor2022measuring}, consisting of a state-preparation layer, a forward random circuit $U$ of $d/2$ cycles, its exact inverse $U^\dagger$, and a readout. The survival probability---the probability of recovering the prepared bitstring---is a direct fidelity estimate for the $d$-cycle sequence $UU^\dagger$ and requires no simulation.

To make the mirror and patched curves comparable at fixed depth, we mirror a \emph{pseudo-patched} circuit constructed to match the $K=3$ patched family gate-for-gate within $\sim2\%$ while remaining connected: the forward circuit cycles through the five $K=3$ boundary-edge sets, omitting in each cycle only the currently active partition's boundary gates, so that removed gates have patched-circuit multiplicity but every coupler is used again under a different partition (Appendix~\ref{app:mirrordetails}). Coherent-error cancellation between the two mirror halves is suppressed by randomized compiling with 64 Pauli-frame randomizations per circuit~\cite{wallman2016noise,hashim2021randomized,proctor2022measuring}, and readout-asymmetry bias is removed by preparing random Hamming-weight-$n/2$ input strings, 10 per circuit, matching the SPAM exposure of the anticoncentrated XEB outputs (Appendix~\ref{app:mirrordetails}). For each mirror point we run 3 independent circuit instances, matching the per-partition instance count of the patched runs.

All fidelity curves and the $10^{6}$-sample run were executed through Qiskit Runtime batch execution, interleaved at the level of individual depth points and measurement-twirled throughout, so that calibration drift affects all curves alike and cross-curve comparisons at fixed depth remain unbiased (Appendix~\ref{app:schedule}).

\section{Estimating the classical cost of sampling}\label{sec:classical}

Every number in this section traces to a stated convention and can be reproduced from the released circuits; the estimator itself is validated against the published advantage networks in Appendix~\ref{app:validation}.

\emph{Task and conventions.}---The task we cost is the one used in Refs.~\cite{morvan2024,kalachev2021,gao2025}: generate $N_s=10^{6}$ samples from a bounded-fidelity approximation to the ideal 78-qubit output distribution---one reproducing the experiment's global fidelity $F$---by frugal rejection sampling from tensor-network amplitude evaluations. In this strategy, drawing $N_s$ samples requires $\kappa\,N_s$ approximate output probabilities with $\kappa\simeq10$ \cite{morvan2024}, and the bounded-fidelity trick makes the work linear in $F$~\cite{kalachev2021}. Writing $\Camp$ for the number of complex multiply--adds in a single amplitude contraction, the estimated work and
runtime are

\begin{equation}
  W \;=\; 8\,\kappa\,N_s\,F\,\Camp
  \quad\text{machine FLOPs},
  \qquad
  t \;=\; \frac{W}{\eta P},
  \label{eq:costmodel}
\end{equation}

with the factor of $8$ converting one single-precision complex multiply--add to machine floating-point operations, and Frontier taken as the reference machine with theoretical peak $P=2.056\times10^{18}$ FLOPS at sustained efficiency $\eta=0.20$~\cite{atchley2023,top500}, following the conventions of Refs.~\cite{morvan2024,gao2025}.

This task is strictly stronger than producing bitstrings that merely match the experimental XEB \emph{score}. Score-level spoofing algorithms attain the latter at far lower cost while carrying parametrically smaller fidelity: samples drawn from the product distribution of two severed halves that reach $\chi=\Fxeb(40)$ would have global fidelity of order $(\chi/2)^{2}\sim10^{-8}$~\cite{gao2024,aaronsongunn2020}. Equation~\eqref{eq:costmodel} is therefore an algorithm-specific estimate for bounded-fidelity sampling of the full circuit, not a lower bound on the cost of matching the reported score; the margin against score spoofing is measured on our own circuit ensemble in Appendix~\ref{app:spoofing}.

Two remarks on conventions. First, $\Camp$ in Eq.~\eqref{eq:costmodel} counts \emph{complex} operations, so all values quoted in this paper are complex FLOPs unless labeled ``machine.'' Second, Eq.~\eqref{eq:costmodel} prices the probabilities as independent single amplitudes. This is the transparent baseline, but it is not what an optimized attack would do, and the corrections do not all point the same way: multi-amplitude and batched sparse-output contraction reuse intermediates across the $\kappa N_s$ evaluations and \emph{lower} the cost~\cite{liu2024,gao2025}, while finite memory forces slicing and recomputation that \emph{raise} it, by amounts that span many orders of magnitude on the published networks~\cite{morvan2024,gao2025}. We therefore quote Eq.~\eqref{eq:costmodel} evaluated with unlimited-memory contraction paths and report the sensitivities separately below.

\emph{Estimator and search protocol.}---Circuits are converted to tensor networks in \textsc{quimb}~\cite{gray2018quimb} and contraction orders are searched with the hyper-optimized protocol of \textsc{cotengra}~\cite{gray2021}: graph-partitioning trials (KaHyPar) under Bayesian hyperparameter search, keeping the best contraction tree found. $\Camp$ is the standard contraction cost of the best tree---the sum over pairwise contractions of the product of participating bond dimensions~\cite{markov2008}---and the reported width $w$ is the base-2 logarithm of the largest intermediate tensor.

Path quality dominates the error of such estimates, so we state the search budgets explicitly. The validation networks of Appendix~\ref{app:validation} were each searched for 5 hours on 128 cores per contraction mode. At small budgets the estimator is badly biased upward---minutes-scale single-core searches overestimate the converged values by two to five orders of magnitude on the same networks---and the bias shrinks with budget until the best-found cost plateaus. The experimental circuits of Table~\ref{tab:depthsweep} were searched with the same protocol under depth-scaled time budgets, from 15 minutes at 8 cycles to 4 hours at 28--40 cycles on 10--16 parallel workers (up to $1.2\times10^{4}$ trials per search). Every depth was searched twice independently, and depths 36 and 40 a third time, for 2 hours, where the independent searches agree to within $0.30$ decades, which we take as the search scatter; since the depth-36 network contains the depth-32 network, the optimal $\Camp(32)$ is in fact bounded above by the depth-36 entry, and we quote each best-found value as is. All quoted costs remain upper bounds on the true optimal contraction cost.

\emph{Depth dependence for the experimental circuits.}---Table~\ref{tab:depthsweep} reports the converged single-amplitude cost for our 78-qubit circuit family at every measured depth.

\begin{table}[t]
\centering
\begin{tabular}{ccccc}
\toprule
cycles & 2q gates & width $w$ & $\log_{10}\Camp$ & $t_{\mathrm{samp}}$ \\
\midrule
 4 &  133 &  1 & --    & -- \\
 8 &  266 & 12 &  5.59 & 8.5\,$\mu$s \\
12 &  399 & 23 &  9.26 & 18\,ms \\
16 &  532 & 34 & 12.70 & 22\,s \\
20 &  665 & 48 & 16.52 & 18.0\,h \\
24 &  798 & 57 & 19.69 & 1.3\,yr \\
28 &  931 & 70 & 23.14 & $1.7\times10^{3}$\,yr \\
32 & 1064 & 78 & 25.70 & $2.8\times10^{5}$\,yr \\
36 & 1197 & 82 & 27.23 & $4.2\times10^{6}$\,yr \\
40 & 1330 & 82 & 27.39 & $2.7\times10^{6}$\,yr \\
\bottomrule
\end{tabular}
\caption{Best-found single-amplitude contraction cost and sampling cost versus depth for the 78-qubit experimental circuits. $\log_{10}\Camp$ is in complex operations; $t_{\mathrm{samp}}$ is the Frontier runtime of Eq.~\eqref{eq:costmodel} for the $10^{6}$-sample ensemble, with $\Fxeb(d)$ taken from the global exponential fit of Fig.~\ref{fig:main}. The sampling cost peaks at the saturation knee, which for these circuits lies near 36 cycles. \emph{Both columns assume unlimited working memory}; memory-constrained contraction is strictly more expensive, by amounts that reach many orders of magnitude on comparable published networks (Sec.~\ref{sec:classical}).}
\label{tab:depthsweep}
\end{table}

Below the knee the cost grows by roughly $10^{3.5}$ per four cycles up to 24 cycles---exponential growth at roughly three width units per cycle---and by $10^{2.5}$ per four cycles from there to 36 cycles, where the width reaches $w=82\approx n$ and the growth stops: between depths 36 and 40 the cost changes by a factor of 1.4. The measured knee position, at 36 cycles, matches the heuristic $d^{\ast}\simeq4\sqrt{n}\approx35.3$ to within one four-cycle step. The coefficient is where the native gate set enters. CZ has operator Schmidt rank 2 and contributes one ebit per gate across a contraction cut, whereas the iSWAP-family gates of Refs.~\cite{arute2019,morvan2024,gao2025} have rank 4 and contribute two~\cite{arute2019,markov2008}; correspondingly, the published networks saturate near $d^{\ast}\simeq3\sqrt{n}$ while our CZ circuits saturate near $d^{\ast}\simeq4\sqrt{n}$. A weaker entangler delays the crossover but does not lower the saturated cost, which is pinned at system size on either gate set.

At the anchor depth this cost is already far beyond direct verification: $\Camp\approx10^{27.4}$ complex operations is $\sim1.5\times10^{3}$ years of the full reference machine for a \emph{single} amplitude, which is why the fidelity at 40 cycles is measured by mirror and patch estimators rather than by full-circuit XEB.

\emph{Sampling cost and sensitivity.}--- Combining Table~\ref{tab:depthsweep} with the measured fidelities gives the headline estimates. At 40 cycles, $\Fxeb(40)\approx1.8\times10^{-4}$ and $\Camp=10^{27.4}$ yield $W\approx3.5\times10^{31}$ machine FLOPs, i.e.\ $t\approx2.7\times10^{6}$ Frontier years at the stated efficiency. At 36 cycles the same formula with the fitted $\Fxeb\approx4.0\times10^{-4}$ and $\Camp=10^{27.2}$ gives $t\approx4.2\times10^{6}$ years, and at 32 cycles ($\Fxeb\approx9.0\times10^{-4}$, $\Camp=10^{25.7}$) $t\approx2.8\times10^{5}$ years: the contraction cost saturates at 36 cycles while the fidelity keeps decaying, so the fidelity-weighted work $F\Camp$ is maximized at the knee, at 36 cycles. Our samples are taken one four-cycle step past it, at 40 cycles, where the task is still far out of reach: $2.7\times10^{6}$ Frontier years under Eq.~\eqref{eq:costmodel}, a figure that a memory limit only raises.

The sensitivities are as follows. (i)~\emph{Fidelity:} $W$ is linear in $F$ within this strategy; the linearity is a property of the bounded-fidelity algorithm~\cite{kalachev2021}. (ii)~\emph{Memory model:} we quote unlimited-memory paths; on the same published network, realistic-memory and unlimited-memory estimates differ by up to twelve orders of magnitude in Ref.~\cite{morvan2024}, so the memory convention dominates every other choice, and quoting the floor is the conservative direction for an advantage claim. With unlimited memory, a state-vector simulation of the 78-qubit, 40-cycle circuit, which gives all $2^{78}$ amplitudes exactly and independently of $F$, would take $1.2\times10^{3}$ Frontier years at the same efficiency, but it needs $2.4$\,YB of memory, eight orders of magnitude more than the $\sim$10\,PB of Frontier~\cite{atchley2023}. A memory limit only raises the cost of contraction: searches capped at Frontier's memory (largest tensor $\le2^{49}$ elements), for single amplitudes and for amplitudes batched~\cite{liu2024}, did not come below $3\times10^{11}$ Frontier years. A Schr\"odinger--Feynman simulation~\cite{aaronsonchen2017,arute2019} across the eight-coupler $34+44$ bisection of the grid (one binary path index per severed CZ, 141\,TB per path) is cheaper, at $2.9\times10^{8}$ years, but equally far out of reach. (iii)~\emph{Path search:} converged-budget searches are upper bounds on the optimal $\Camp$ with residual scatter $\lesssim0.30$ decades at saturation. (iv)~\emph{Algorithm:} Eq.~\eqref{eq:costmodel} does not price multi-amplitude reuse~\cite{liu2024}, approximate contraction~\cite{graychan2024}, finite-fidelity matrix-product simulation~\cite{ayral2023}, or spoofing strategies that target the score rather than the distribution~\cite{gao2024,aharonov2023}; any of these can reduce the effective cost~\cite{liu2021,pan2022,zhao2025}.

\section{Discussion}\label{sec:discussion}

We have run random-circuit sampling at the scale of the published advantage experiments on a publicly available QPU accessible through a cloud, using the code that we release with this paper. The full experiment---fidelity estimators at depths from 4 to 40 cycles plus the $10^{6}$-sample run of the 78-qubit, 40-cycle circuit---consumed about 9 minutes of QPU execution time over two executions, of which each sampling run took 22\,s (Appendix~\ref{app:schedule}). The two estimators, one simulation-free (the mirror benchmark) and one built on exact patch simulation, agree at every depth from 20 to 40 cycles.

The directly supported point is the 78-qubit, 40-cycle experiment, one four-cycle step past the knee of the contraction cost at 36 cycles; the data show that both estimators keep following the same exponential through the knee, and at 40 cycles, under the cost model of Sec.~\ref{sec:classical}, the classical task is slightly cheaper than at the knee but still takes more than a million Frontier years. Score spoofing and noise-induced classical algorithms limit what any finite-fidelity RCS experiment can claim, and simulation methods keep getting faster~\cite{huang2020,liu2021,pan2022,zhao2025,aaronsongunn2020,barak2021,gao2024,aharonov2023}; this is why we quote unlimited-memory cost floors rather than memory-constrained headline numbers. For the severing attack family specifically, Appendix~\ref{app:spoofing} measures the attainable scores on our own ensemble: no branch reaches the 40-cycle score, and within this family no strategy attains the reported score more cheaply than sampling the circuit itself. The analysis also leaves a concrete lesson for future placements: optimize graph expansion alongside gate fidelity.

\section*{Acknowledgments}
We thank Petar Jurcevic for many fruitful discussions and acknowledge the use of IBM Quantum Credits via the IBM Quantum Startups Program for this work. The views expressed are those of the authors and do not reflect the official policy or position of IBM or the IBM Quantum Platform team.

\section*{Data and code availability}
The data and code that support the findings of this work are openly available at \url{https://github.com/BlueQubitDev/rcs-nighthawk}. The repository contains all executed circuits---the mirror, patched, and full-circuit instances in QPY and OpenQASM~3 format, together with the qubit layout, the coupler four-coloring, the patch partitions, and the measured bitstrings, including the $10^{6}$ samples of the 40-cycle circuit. The code for circuit generation, calibration-aware placement, fidelity estimation, and contraction-cost estimation is also available.

\appendix
\section{Placement details}\label{app:placement}

For every possible offset $(\delta r,\delta c)$ of the logical $9\times9$ grid inside the $12\times 10$ physical lattice ($4\times 2 = 8$ candidate rectangles) we compute a log-fidelity proxy
\begin{equation}
\begin{split}
\log F(\delta r, \delta c)\;=\;
  &\sum_{q}\log\!\bigl(1 - e^{\mathrm{ro}}_{q}\bigr)
+ \sum_{q}\log\!\bigl(1 - e^{\mathrm{sx}}_{q}\bigr) \\
+ &\sum_{\langle q,q'\rangle}\log\!\bigl(1 - e^{\mathrm{cz}}_{qq'}\bigr),
\end{split}
\label{eq:placement-proxy}
\end{equation}
where $e^{\mathrm{ro}}$, $e^{\mathrm{sx}}$ and $e^{\mathrm{cz}}$ are the reported readout, SX and CZ error rates and the sums run over the qubits and couplers of the candidate rectangle that pass the filters below. Within a candidate rectangle, qubits whose readout error exceeds $4\%$ are dropped, couplers whose CZ error exceeds $2\%$ are pruned, and any qubit left with no usable coupler is dropped as well. Among all candidates whose proxy fidelity is within a factor $e^{-0.1}\approx0.9$ of the best one (a slack of $0.1$ in $\log F$), we select the one retaining the most qubits. This is a coarser, purpose-built variant of noise-aware layout selection in the spirit of \textit{mapomatic}~\cite{nation2023suppressing}: because the circuit family is translation-invariant on the grid, only a few rigid placements need to be scored.

\section{Patch generation}\label{app:patchgen}

Patch partitions are produced by a randomized combinatorial optimizer operating on the 133-edge coupling graph, using boundary local search that moves single qubits between patches, subject to two constraints---every patch remains internally connected, and patch sizes stay balanced to within one qubit---while minimizing the number of cut edges. The optimizer returns the set of distinct optimal partitions found, ranked by cut size. For our experiment this yields for $K=3$: $26+26+26$ qubits with 16--21 cut edges; for $K=4$: $19+19+20+20$ qubits with 19--26 cut edges. Five partitions are retained at each $K$.

These partitions are optimized for fidelity estimation (balanced, simulable patches), not for minimal boundary, and an adversary is not bound by the balance constraint. A randomized min-cut search over the same graph (Appendix~\ref{app:spoofing}) gives, for cuts with both sides larger than 33 qubits, a minimum of 8 cut edges (a $34+44$ partition); relaxing the balance requirement further, blocks of 6 to 14 qubits detach on 5 edges, a 4-qubit block on 3, a 3-qubit block on 2, and the graph carries eight degree-2 qubits that detach on 2. The gap between the balanced estimator partitions (16--26 cut edges) and these adversarial cuts (2--8) is what makes the score-spoofing analysis of Appendix~\ref{app:spoofing} necessary: the estimator cuts are deliberately expensive to sever, while the cuts an attacker would actually choose are up to an order of magnitude cheaper.

\section{Mirror construction details}\label{app:mirrordetails}
\begin{table*}[t]
\centering
\begin{tabular}{lcccccccc}
\toprule
 & & & & & \multicolumn{3}{c}{$\log_{10}\Camp$} & \\
\cmidrule(lr){6-8}
experiment & $n$ & cycles & 2q gate & $\Fxeb$ & published & ours & $\Delta$ & $t_{\mathrm{samp}}$ [Eq.~\eqref{eq:costmodel}] \\
\midrule
Sycamore-53~\cite{arute2019}   & 53 & 20 & fSim       & $2\times10^{-3}$   & 17.78 & 18.40 & $+0.62$ & 11\,d \\
Zuchongzhi-56~\cite{wu2021}    & 56 & 20 & iSWAP-like & $6\times10^{-4}$   & 19.78 & 19.44 & $-0.34$ & 37\,d \\
Zuchongzhi-60~\cite{zhu2022}   & 60 & 24 & iSWAP-like & $3\times10^{-4}$   & 21.00 & 20.85 & $-0.15$ & 1.3\,yr \\
Sycamore-70~\cite{morvan2024}  & 70 & 24 & fSim       & $2\times10^{-3}$   & 23.70 & 23.00 & $-0.70$ & $1.2\times10^{3}$\,yr \\
Sycamore-67~\cite{morvan2024}  & 67 & 32 & fSim       & $1\times10^{-3}$   & 23.30 & 23.31 & $+0.01$ & $1.3\times10^{3}$\,yr \\
Zuchongzhi-3.0~\cite{gao2025}  & 83 (of 105) & 32 & iSWAP-like & $2.5\times10^{-4}$ & --- & $29.1^{\,a}$ & --- & $1.7\times10^{8}$\,yr \\
\addlinespace[2pt]
Quantinuum H2~\cite{decross2025}       & 56 & 20 & RZZ$^{\,b}$ & $0.17$  & --- & 19.8 & --- & 57\,yr \\
Quantinuum Helios~\cite{ransford2026}  & 98 & 26 & RZZ$^{\,b}$ & $0.045$ & --- & 32.9 & --- & $2\times10^{14}$\,yr \\
\addlinespace[2pt]
IBM Nighthawk r2 (this work)  & 78 & 40 & CZ & $1.8\times10^{-4}$ & --- & 27.4 & --- & $2.7\times10^{6}$\,yr \\
\bottomrule
\end{tabular}
\caption{The forward-RCS record under one convention, and the validation of the estimator that puts it there. $\log_{10}\Camp$ is in complex operations for one exact, noiseless output amplitude with unlimited memory (arXiv convention). For the five circuits whose exact tensor networks are released~\cite{rcstnsa}, ``published'' is the value tabulated in Ref.~\cite{morvan2024} and ``ours'' the best found by our searches (5\,h $\times$ 128 cores per contraction mode); agreement is within $0.7$ decades on all five and within $0.01$ on the deepest. $\Fxeb$ for those rows is as tabulated in Ref.~\cite{morvan2024}. $t_{\mathrm{samp}}$ evaluates Eq.~\eqref{eq:costmodel} at $N_s=10^{6}$ with $\kappa=10$ on Frontier. $^{a}$Geometry-approximate reconstruction; the exact qubit selection is not public. $^{b}$All-to-all connectivity: each layer applies $n/2$ $R_{ZZ}(\pi/2)$ gates on a random pairing. $\Camp$ is computed on the exact released circuits~\cite{cqcl_rcs_data} with the same search protocol: the median over five instances for H2 (spread $0.16$ decades, and within $0.23$ decades of an independent simulated-annealing search~\cite{rcstnsa}), the single released challenge circuit for Helios. The fidelities are mirror-benchmarking estimates, as ours are: for H2, the return probability of the released mirror circuits evaluated with the released analysis code; for Helios, the published gate-counting fit at depth 26.}
\label{tab:record}
\end{table*}

\emph{Pseudo-patched forward circuit.}---A mirror of the full circuit would contain more two-qubit gates per cycle than the patched circuits it is compared against (the patched variants lack the boundary gates), so the two benchmarks would not be exactly comparable at fixed depth---their per-cycle error budgets would differ. The forward circuit therefore cycles through the same five $K=3$ boundary-edge sets used for randomized patching, advancing to the next partition every cycle; in each cycle, the CZ gates belonging to the currently active partition's boundary are omitted from that cycle's matching. Every removed gate is thus a genuine patch-boundary gate of one of the partitions and the number of omitted gates per four-cycle period matches a $K=3$ patched circuit. But because the active partition rotates, a coupler severed in one cycle is used again at its next appearance under a different partition, so the interaction graph of the forward circuit remains connected. The construction sits between the ``patched'' and ``elided'' verification circuits~\cite{arute2019,morvan2024}---gates are removed with patched-circuit multiplicity, but the circuit stays connected as in the elided variant---while additionally being mirrored. As a result the mirror curve and the patched curves can be plotted against either depth (cycles) or two-qubit gate count: the ``cycle picture'' and ``gate-count picture'' coincide on the $x$-axis by construction.

\emph{Gate twirling.}---Mirror circuits are vulnerable to coherent-error cancellation: systematic control errors in $U^\dagger$ can partially undo those of $U$, biasing survival probabilities optimistically relative to the error accumulated by a non-mirrored circuit. We suppress this with gate twirling (randomized compiling)~\cite{wallman2016noise,hashim2021randomized,proctor2022measuring}: each executed randomization dresses the two-qubit gate layers with independently sampled Pauli frames, with 64 randomizations per circuit, converting coherent errors into stochastic Pauli channels whose contributions add rather than cancel between the two mirror halves.

\emph{Input randomization.}---Readout on superconducting devices is asymmetric: confusion-matrix calibrations on Nighthawk r2 QPU give a median $P(1\!\to\!0)$ misassignment of $\sim\!1.0\%$ against $P(0\!\to\!1)\sim\!0.3\%$. A mirror benchmark prepared on the all-zeros state would therefore see only the smaller error branch and would be biased high relative to the XEB curves, whose anticoncentrated output strings have a higher average Hamming weight and thus experience the symmetrized readout error $(e_{0}+e_{1})/2$ per qubit. To make the comparison fair, we randomize the input to mirror circuits: the preparation layer applies $\mathrm{RX}(\pi)$ on a uniformly random half of the qubits, giving random Hamming-weight-$n/2$ bitstrings. Each mirror circuit is executed on 10 random input strings with an equal share of the shot budget, and the survival probability is averaged over inputs. This makes the SPAM contribution of the mirror benchmark closer to that of the patched XEB curves.

\section{Execution schedule, drift mitigation, and measurement twirling}
\label{app:schedule}

The full experiment---the mirror benchmark at 4--40 cycles, the $K=3$ and $K=4$ patched families at 12--40 cycles, and the $10^{6}$-sample supremacy run of the unpatched 40-cycle circuit---was executed twice on \textit{ibm\_phoenix}, with identical circuits on the same qubits, through Qiskit Runtime batch execution, with $1.6\times10^{7}$ shots in total and about 9 minutes of QPU execution time at the backend's default shot repetition delay of $1\,\mu$s; each $10^{6}$-sample supremacy run took 22\,s. Counts of each circuit are summed over the two executions.

Drift in gate and readout calibrations over minutes motivates the job order: if each curve were collected end-to-end before starting the next, slow drift would give every curve a different effective error rate and break the cross-curve comparisons at fixed depth that our analysis relies on. We therefore interleave the curves at the level of individual data points: the execution iterates over depths in ascending order and, at each depth, submits back-to-back one job per curve that includes that depth---first the mirror, then $K=3$ and $K=4$. A patched job carries all 15 circuit instances of a data point (five partitions, three instances each) and a mirror job its $3\times10$ instance--input combinations, so all curves sample any given calibration epoch at the same depths. Hardware drift then shows up as correlated fluctuations that make the individual curves somewhat bumpier, but relative comparisons between curves at fixed depth remain unbiased. Per-instance shot budgets grow with depth to track the shrinking fidelity signal, from $2\times10^{3}$ to $9\times10^{4}$ shots per patched circuit and from $6\times10^{4}$ to $6\times10^{5}$ per mirror instance, summed over both executions. Fidelities are combined across instances using an inverse-variance weighted average.

For the same reason we enable measurement twirling on every job in the session. This symmetrizes the readout, so that slow drift in readout error cannot bias one curve relative to another depending on when it was collected, and it makes the readout-error contribution identical in kind across the mirror and XEB benchmarks. Gate twirling is applied to the mirror circuits only (Appendix~\ref{app:mirrordetails}), since the XEB estimator is insensitive to the coherent-cancellation effect that motivates twirling in mirrored circuits.

\section{Validation of the contraction-cost estimator}\label{app:validation}

The estimator was validated on the tensor-network graph files released with Google's contraction-order optimizer~\cite{rcstnsa}, which specify the exact networks behind the single-amplitude column of Table~1 of Ref.~\cite{morvan2024}. Table~\ref{tab:record} compares our best-found $\Camp$ against the published values for all five circuits.

Agreement is within $0.7$ decades on all five networks and within $0.01$ decades on the deepest (Sycamore-67 at 32 cycles), which is the circuit most comparable in depth to our experiment. As an end-to-end check at larger scale, we also rebuilt a geometry-approximate Zuchongzhi-3.0 network (83 qubits, 32 cycles, iSWAP-type gates in the published ABCDCDBA schedule) and evaluated Eq.~\eqref{eq:costmodel} at their reported $\Fxeb=2.5\times10^{-4}$: the resulting runtime is within a factor of four of the unlimited-memory estimate of Ref.~\cite{gao2025} ($5.7\times10^{7}$ Frontier years), the appropriate comparison row for our memory convention. Because the exact qubit selection of that device is not public, we regard this as a consistency check rather than a reproduction.

\section{Severing attacks and placement on the weak-noise side}
\label{app:spoofing}

\begin{table}[t]
\centering\small
\setlength{\tabcolsep}{4pt}
\begin{tabular}{lccc}
\toprule
cut & $b$ & $\chi_{\mathrm{spoof}}(40)$ & margin \\
\midrule
balanced $34{+}44$ & 8 & $10^{-29}$--$10^{-20}$ & $6^{\ddagger}$ \\
6- to 14-qubit blocks & 5 & $10^{-18}$--$10^{-12}$ & $6^{\ddagger}$ \\
block qubit$^{\dagger}$ & 2 & $1.1(0.3)\times10^{-4}$ & $1.6$ \\
3-qubit block & 2 & $\approx2\times10^{-5}$ & $10$ \\
adjacent pairs$^{\ast}$ & 2--4 & $<6.8\times10^{-5}$ & $2.6$ \\
singletons$^{\ast}$ & 2 & $<4.7\times10^{-5}$ & $3.8$ \\
\bottomrule
\end{tabular}
\caption{Severing-attack branches against the 40-cycle, $n=78$ experiment. ``margin'' is the target $\Fxeb(40)=1.8\times10^{-4}$ divided by the attainable score; $b$ is the number of severed couplers. Rows marked $^{\ast}$ are one-sided (only the small side simulated, uniform bits elsewhere), which costs the attacker essentially nothing, and give $2\sigma$ upper bounds of ensemble means that are consistent with zero; the others sample both severed parts. Scores are ensemble means from exact inner products on regions of the realized coupler graph ($1{,}680$--$20{,}040$ instances per region, four regions); the 3-qubit-block row is the 40-cycle value of its decay resolved up to 28 cycles, and the top two rows are boundary terms extrapolated from the fitted rates $c=0.44$--$0.55$, $K=1.2$--$38$. $^{\ddagger}$The margin of a wide cut is not set by its boundary term but by the slow modes of the low-degree qubits inside its sides, $\lesssim3\times10^{-5}$. No branch reaches the experimental score. The mode marked $^{\dagger}$, the corner qubit of the 3-qubit block severed alone, comes within a factor of two of it; it requires exact samples of the 77-qubit remainder, at $\approx3\times10^{3}$ times the cost of the baseline of Eq.~\eqref{eq:costmodel}.}
\label{tab:spoof}
\end{table}

The linear XEB score can be attained by a dishonest classical spoofer without sampling from any bounded-fidelity approximation of the circuit. The relevant family is the severing attack~\cite{gao2024,barak2021,aaronsongunn2020}: partition the qubits into $(A,\bar A)$, delete the CZ gates crossing the cut, sample the severed parts independently, and submit the product samples. Whether such attacks succeed at a given operating point is a phase question---the measured score contains a global sector decaying with the full error budget and subsystem sectors decaying with a subsystem's error budget plus a boundary penalty, and the competition between the two is the noise-induced transition of Refs.~\cite{morvan2024,ware2023}. Our ensemble, CZ entanglers with Haar-random SU(2) rotations, is the family for which severing is most effective: Ref.~\cite{gao2024} reaches $\chi\simeq0.024$ for a CZ ensemble on the Sycamore-53 geometry, versus $8\%$ of the experimental score for the actual fSim ensemble. We therefore quantify the margin on our own circuits rather than by transfer from published geometries. The attack scores below are properties of the ideal circuits and the coupler graph; device data enters only through the fidelity fit that sets the comparison target.

\emph{Setup.}---For a cut severing $b$ couplers, each fired once per four-cycle sweep, the attack score decays as $\chi(d)\simeq K\,c^{\,(b/4)d}$, with $c$ the score fraction retained per severed CZ, while the global signal decays as $\Fxeb(d)\simeq0.566\times10^{-0.0875d}$, a factor $0.818$ per cycle [the fit of Fig.~\ref{fig:main}]. At the rate level a subsystem sector outruns the global signal if its per-cycle attenuation is the weaker one, $c^{\,b/4}>1-\epsilon_q(n-|A|)$, with $\epsilon_q\simeq2.6\times10^{-3}$ the per-qubit per-cycle error; at finite depth the prefactors compete with the accumulated decays, and for small $|A|$ the uniform-$c$ approximation itself fails (the effective decay of few-qubit marginals is governed by a size-dependent gap~\cite{gao2024}). Rate margins alone therefore settle nothing for the small-subsystem modes; both $c$ and $K$ have to be measured, cut by cut.

\emph{Adversarial cuts of the realized coupler graph.}---The partitions of Appendix~\ref{app:patchgen} minimize cut size under a patch-balance constraint appropriate for fidelity estimation; an adversary is not so constrained. A randomized min-cut search over the realized 133-edge graph finds a balanced $34+44$ cut severing $b=8$ couplers, blocks of 6 to 14 qubits attached by $b=5$, a 4-qubit block attached by $b=3$, a 3-qubit block (qubits 68, 76, 77) attached by only $b=2$, and eight degree-2 qubits ($b=2$ singleton modes), three of them the qubits of that block. All are low-expansion motifs, left because the placement optimizer sees only calibration data (Appendix~\ref{app:placement}).

\emph{Exact attack scores on ensemble instances.}---For targets restricted to $m\le24$-qubit regions of the coupler graph the attack score is an exact inner product of the target probability vector with the severed product distribution, free of sampling noise. Two practical points govern the measurement. First, the single-instance score of a small-$|A|$ mode fluctuates by $\sim2^{-(m-|A|)/2}\approx10^{-3}$ from circuit to circuit, burying the deep-circuit mean; we therefore average $1{,}680$--$20{,}040$ independently drawn instances of the ensemble (Haar SU(2) layers over the experiment's color classes) in each of four rectangular regions of 19--20 qubits centered on the weak motifs above, resolving means at the $3\times10^{-5}$ level. The regions are chosen so that no qubit is left attached by a single coupler: such pieces relax slowly, keep the collision ratio of the region above its Porter--Thomas value, and inflate every two-sided score. Second, the fitted decays of the $b=2$ modes are boundary-local: the retained fraction is $c=0.44$--$0.55$ per severed CZ with prefactors $K=1.2$--$38$, and at $n=78$ the per-instance speckle is $\sim2^{-(n-|A|)/2}\lesssim10^{-11}$, so the ensemble means measured here are the operative scores at full size.

The result separates the branches by what they cost the attacker (Table~\ref{tab:spoof}). The one-sided modes, which are essentially free to run, are score-dead at the 40-cycle anchor, where the target is $\Fxeb(40)=1.8\times10^{-4}$: every one of them is consistent with zero, the singletons within $4.7\times10^{-5}$ and the adjacent pairs within $6.8\times10^{-5}$, and the one-sided mode of the whole 3-qubit block, whose decay is resolved up to 24 cycles, extrapolates to $7\times10^{-6}$. The two-sided modes inside the 3-qubit block are not. Each of its qubits is itself attached by two couplers, and severing one of them alone leaves its neighbors in the block hanging on a single coupler in the severed remainder; the score of the two severed parts then decays by a factor $0.34$ per four cycles, against $0.45$ for the fidelity. For the corner qubit of the block it is $4.8(0.4)\times10^{-4}$ at 36 cycles and $1.1(0.3)\times10^{-4}$ at 40 cycles, that is $1.2(1)$ and $0.6(2)$ times the experimental fidelity at those depths: above it up to 36 cycles, and within a factor of two of the target at the anchor. For the other two qubits it is $0.5(0.3)\times10^{-4}$ and $0.9(0.3)\times10^{-4}$ at 40 cycles, and for the block severed as a whole $\approx2\times10^{-5}$. Extrapolating the fitted rates to the wider cuts that an adversary would need for a balanced partition gives boundary terms $\chi(40)\sim10^{-18}$--$10^{-12}$ for the 6- to 14-qubit blocks ($b=5$) and $10^{-29}$--$10^{-20}$ for the balanced $34+44$ cut ($b=8$). A two-sided attack on such a cut also reproduces the slow modes of the low-degree qubits that lie inside its sides, which we find to contribute at most half the collision-ratio excess of the surrounding region, $\lesssim3\times10^{-5}$ at 40 cycles for the region around the 3-qubit block; its total score is set by those modes, a factor of 6 below target, not by its own boundary term.

\emph{Margin at the operating point.}---No branch reaches the experimental score at the anchor, but the strongest comes within a factor of two of it, too close to exclude on score alone. Such a branch must, however, still produce the samples. The severed complement of a single qubit is a 77-qubit RCS circuit whose saturated single-amplitude cost is $\approx2^{-1}\Camp$. The scores above assume that it is sampled exactly, which by Eq.~\eqref{eq:costmodel} costs $2^{-1}/\Fxeb(40)\approx3\times10^{3}$ times the full-circuit baseline of Sec.~\ref{sec:classical}; sampling it at fidelity $f$ lowers cost and score together by $f$. Within the severing family, then, no strategy attains the reported 40-cycle score at a cost below the faithful-sampling baseline. The attainable severed scores are set by the ideal circuits, so raising fidelity raises the bar a spoofer must clear without giving the spoofer anything, and would restore a margin on score alone. At the rate level the conclusion holds with margin for \emph{every} cut the graph admits, including the singleton modes: the boundary bill exceeds the noise saved by $1.5$--$2.0 \times$ for a degree-2 singleton and by $11$--$14\times$ for the balanced cut, so the global signal gains on every severed score as depth increases. Known top-$k$ postprocessing improves severed scores by $O(1)$ factors~\cite{gao2024} and does not change the conclusion. These are attack-class-relative statements, on the same footing as the corresponding analyses of Refs.~\cite{morvan2024,gao2025}.

\emph{Anticoncentration.}---The same computation certifies the first crossover. The Porter--Thomas collision ratio $2^{m}\sum_x p(x)^2-1$ reaches $\approx1.01$ by 24 cycles, around $1.00$ by 32 cycles and remains there at 36 and 40 cycles, placing the anchor depth well inside the anticoncentrated regime and away from the shallow-circuit spoofing window of Refs.~\cite{barak2021,dalzell2022}.

\emph{Implication for placement.}---The weak motifs above---blocks of 6 to 14 qubits on five edges, a 4-qubit block on three, a 3-qubit block on two, eight degree-2 qubits---are all artifacts of a placement objective that optimizes calibration data alone (Appendix~\ref{app:placement}). A placement that additionally penalizes low-conductance cuts, trading a small amount of median gate fidelity for graph expansion, would remove them and widen the margin further at essentially no cost in circuit fidelity; we regard this as a design rule for future experiments. The scaling caveat is set by geometry: at fixed per-gate error on a fixed-degree planar lattice the boundary bill of the best cut grows only as $\sqrt{n}$ while the global error budget grows as $n$, so at constant $\epsilon_q$ sufficiently large systems cross to the strong-noise side~\cite{ware2023,morvan2024,aharonov2023}; improved fidelity or connectivity moves the crossing outward.

\bibliography{refs}

\end{document}